
\headline={\ifnum\pageno=1\firstheadline\else
\ifodd\pageno\rightheadline \else\leftheadline\fi\fi}
\def\firstheadline{\hfil}
\def\rightheadline{\hfil}
\def\leftheadline{\hfil}
	\footline={\ifnum\pageno=1\firstfootline\else\otherfootline\fi}
\def\firstfootline{\rm\hss\folio\hss}
\def\otherfootline{\hfil}
\def\arcsper{\ifmmode \rlap.{'' }\else $\rlap{.}'' $\fi}
\def\ts{\thinspace}

\font\twelvebf=cmbx10 scaled\magstep 1
\font\twelverm=cmr10 scaled\magstep 1
\font\twelveit=cmti10 scaled\magstep 1

\font\tenbf=cmbx10
\font\tenrm=cmr10
\font\tenit=cmti10

\font\ninerm=cmr9

\parindent=1.5pc
\hsize=6.0truein
\vsize=8.5truein
\nopagenumbers

\centerline{\tenbf SURVEYS FOR Z $>$ 1 FIELD GALAXIES}
\baselineskip=22pt

\vglue 0.8cm
\centerline{\tenrm DAVID CRAMPTON$^{*}$ and  C.L. MORBEY}
\baselineskip=13pt
\centerline{\tenit Dominion Astrophysical Observatory, National Research
Council of Canada}
\baselineskip=12pt
\centerline{\tenit  Victoria V8X 4M6, Canada}
\vglue 0.3cm
\centerline{\tenrm O. LE FEVRE$^{*}$, F. HAMMER$^{*}$ and L. TRESSE$^{*}$}
\baselineskip=13pt
\centerline{\tenit D.A.E.C, Observatoire de Meudon, 92195 Meudon, France}
\vglue 0.3cm
\centerline{\tenrm S. J. LILLY$^{*}$ and D.J. SCHADE}
\baselineskip=13pt
\centerline{\tenit Department of Astronomy, University of Toronto, Toronto,
M5S 1A7, Canada}
\vglue 0.8cm
\centerline{\tenrm ABSTRACT}
\vglue 0.3cm
{\rightskip=3pc
 \leftskip=3pc
 \tenrm\baselineskip=12pt\noindent

In the course of the comprehensive CFRS redshift survey to
$ I_{AB}$= 22.5 (see Lilly
{\twelveit et al}, these proceedings) with the CFHT MOS/SIS spectrograph,
we have observed $\sim$25 field galaxies with z $>$ 1. However, it is clear
that
almost an equal number, predominantly red galaxies, have likely been missed
largely because of limitations in the observed wavelength range. The
properties of the observed galaxies and the reasons for the presumed
incompleteness are briefly discussed in this contribution, and strategies
for improving our methods of observing high redshift galaxies are explored.
In particular, it is demonstrated that sky subtraction to the required
limiting magnitude is possible even with very low spectral resolution,
and that new IR arrays and new optical designs for multi-object spectrographs
 will allow
us to easily reach field galaxies at z $>$ 1.

\vglue 0.6cm}

\vfil
\twelverm\baselineskip=14pt
\leftline{\twelvebf 1. Introduction}
\vglue 0.3cm

\vglue 1pt
Spectra of $\sim$1050 objects  to a limiting isophotal
$I_{AB}$ = 22.5 were obtained during the course of the Canada-France Redshift
Survey
(CFRS). As discussed by Lilly {\twelveit et al}, spectra were taken
of all objects to this limiting magnitude, so no observational bias is present
against compact galaxies or AGNs, for example. However, the spectroscopic
wavelength
range was limited to 4200 -- 8600\AA  \ts in order that three tiers of
spectra could be observed simultaneously. All spectra were reduced
independently
by three members of the team and reliability grades or ``notes" were assigned,
for which quite accurate assessments of reliability were derived from repeat
observations of the objects. In addition, comparison of the spectra of galaxies
 at lower
redshift which had strong [OII] and/or H$\alpha $ lines with those of
objects at higher z
which only displayed a single strong  emission line in their spectra (due
to the fact that most features had been redshifted out
 of the observed range) allowed
us to develop a technique to assign unambiguous redshifts for these objects
(Lilly {\twelveit et al}).
-----------------------------------------

\noindent
$^{*}$ {\ninerm Visiting Observer with the Canada-France-Hawaii Telescope,
operated by
the NRC of Canada, the CNRS of France and the University of Hawaii}
\eject

This technique fails for single-line galaxies at
 the very highest redshifts because
of the limited useful spectral range observable, but the lack of other
features and slope of the continua make it highly likely that the assigned
redshifts for these galaxies is also correct. The CFRS survey is in the final
stages of analysis, and a few details may change, but the overall conclusions
discussed here will not.
\vglue 0.6cm
\leftline{\twelvebf 2. Observing z $ > $ 1 Galaxies}
\vglue 0.3cm
\leftline{\twelveit 2.1. Sky Subtraction}
\vglue 1pt
As the redshift of galaxies increases to z $\sim$ 1,
the standard features used to determine redshifts
in low resolution spectra are redshifted into a region of markedly
increased emission from the night sky (one reason why the CFRS spectra
were limited to z$<$8600\AA). Although a few features remain visible (e.g.,
3727[OII] and Ca H and K), the increased noise at longer
wavelengths and the difficulty of
distinguishing ``breaks" from features due to poorly-subtracted night sky
emission means that deriving reliable redshifts for
 ``non-emission line" galaxies at such high redshifts
is virtually impossible for such spectra, in agreement with the conclusions
regarding incompleteness in the previous section.

\vskip 8truecm
{\tenrm Fig. 1. A CFHT MOS/SIS spectrum of a galaxy at z $\sim$0.4, with the
night sky superimposed (dashed line). Note that although the sky emission
increases strongly towards redder wavelengths, the effectiveness of
the sky subtraction is poorer at both blue and red wavelengths,
and hence it is largely a function of the quantum efficiency
of the CCD detector}
\vglue 0.3cm
The conventional wisdom is that this virtual impossibility of deriving
redshifts for z $> $ 1 objects is due to the rapidly increasing brightness
of the night sky at longer wavelengths, combined, of course, with the
decreasing brightness of the objects at increasing z. Looking at figure 1,
the origin of this
belief is easily understood, since the sky-to-object contribution increases
dramatically as a function of wavelength. However, a closer look at the
sky-subtracted spectrum of the  object shows that the increase in the noise
is not significantly worse in the region near 8600\AA  \ts where the
sky has become very bright (this spectrum is not one
of the standard CFRS spectra, but extends to the long wavelength limit of
the CCD). In fact, if the principal emission lines from the object
are removed, and the residual spectrum displayed in counts instead of
flux, it is obvious that sky subtraction is not the major limiting
factor at all, but rather the CCD response. It should perhaps be emphasized
that the original spectrum only had a resolution of $\sim$40\AA,
 or R $\sim$ 200 at
8000\AA, but the sampling is higher than the nominal 2.5 pixels which has
become the default standard in astronomy; there were 5 pixels
per spectral resolution element and the spatial sampling was 0\arcsper 31 per
pixel. The conclusion, which to some extent is based on our whole CFRS dataset,
is that sky subtraction is not the chief limiting factor (as long as
other instrumental factors have been taken care of, particularly flexure,
orthogonality of the slits to the dispersion direction, CCD alignment and
accuracy of the slitlet manufacture). As mentioned above, the declining
response of the quantum efficiency of the CCD is a major factor.
\vglue 0.3cm
\leftline{\twelveit 2.2. New Technical Developments}
\vglue 1pt
Although there has been some success in improving the red quantum efficiency
of CCDs (eg., Schemp$^1$) their physical properties precludes any
response beyond 1.08 $\mu$. Fortunately there are spectacular advances
in the development of 1024x1024 near-IR arrays; the ``Aladdin''$^2$
InSb array should have a quantum efficiency $>$ 85\%
from 0.7$\mu$ to $>$ 2.4$\mu$, and the ``Hawaii''$^3$ HgCdTe array
is purported to have a quantum efficiency $>$ 50\% over the same range.
New, very transparent glasses (e.g. Schott ULTRAN)
have also allowed optical designs to be
extended from the traditional ``visible'' light region into the near-IR
as well. A recent design for an imaging spectrograph for CFHT
yields superb optical images at all wavelengths from 0.4$\mu$ to 2.2$\mu$
with transmission, including reflection losses, of $\sim$80\% over the
entire range. The combination of these new optical designs and new near-IR
arrays will allow spectra of high redshift galaxies to be obtained with
multi-object (uncooled)
spectrographs  in
the accessible ground-based windows from 0.7 - 2.0$\mu$. For redshift
surveys where the redshifts are {\twelveit a priori} unknown, wide wavelength
regions
are essential so at least two prominent spectral features can be observed.
 Even if the wavelength range is limited to $<$ 1.4$\mu$,
redshifts can easily be determined up to z $\sim$ 2 (see section 4).

\vglue 0.6cm
\leftline{\twelvebf 3. CFRS z $ > $ 1 galaxies}
\vglue 0.4cm
There are approximately 23 galaxies with z  $ > $ 1 in the CFRS catalog,
yet the observed luminosity function for  0.5 $<$ z $<$ 0.75  (where we
have very good data and statistics) indicates that at least twice that
number should have been observed. Examination of the
restframe colour--absolute magnitude diagrams at increasing redshift
(figure 2) demonstrates that the galaxies with colours redder than Sbc are
missing at  z $ > 1$. Of course, this could be simply be an effect of
galaxy evolution, but the facts that many of our failures (i.e., objects
for which no redshift was derived) are located in the area of the
(B-I) -- (I-K) two colour diagram where red high z objects are expected to lie,
and that it is difficult to identify non-emission line objects
at z $ > $ 1 with our limited spectral range, lead us to the conclusion that
there are probably $\sim$50 galaxies in our sample which have z $ > $ 1.

\vskip 8truecm
{\tenrm Fig. 2. Derived B absolute magnitudes versus restframe colours
for galaxies in different redshift slices. Galaxies redder than local
Sbc galaxies (to the right of the dotted vertical line in each panel)
were not found at the very highest redshifts, most likely for instrumental
reasons}
\vskip 0.3truecm
Given that the new IR array detectors combined with new optical designs
will allow spectra to be obtained in the required wavelength interval,
are the high redshift galaxies observable in sufficient numbers
with current telescopes?
Since in our large
CFRS survey of $\sim$750 galaxies  only
$\sim$50 have z $>$ 1, it might appear at first sight rather daunting.
The CFRS sample was selected in a completely unbiased way, so
if the subsequent biases are acceptable,
size and colour criteria or radio or X-ray emission could be used to
enhance the probability of selecting z $>$ 1 galaxies. Unbiased samples
could also be selected at redder wavelengths.
For example, assuming relatively
mild evolution (evolution reaching 1.3 mag for z $ > $ 1.3) the redshift
distribution for I and J selected samples are compared in figure 3 to
our observed CFRS I $< $ 22 data. Selection of J $<$ 22 (or H $<$ 21)
galaxies gives a relatively large fraction  ($>$ 25 \%) of z $> $ 1 galaxies
compared to
the CFRS sample, or even a fainter I selected sample.

Are galaxies at these faint red magnitudes realistically observable? Our
CFRS data, somewhat inadvertently, demonstrate that it is feasible. The
CFRS galaxies were selected on the basis of isophotal magnitudes, and
crowding and confusion at the faint limit led us to observe some
galaxies fainter than our survey limit. The mean $I_{AB}$ in a 3'' aperture of
the 25 z$>$ 1
galaxies (see examples in figure 4) is 22.6, and the faintest is 24.4. Hence
it is possible even with our present instrumentation and, with
the advent of higher quantum efficiency (thinned) CCDs
 at CFHT, observations
of such galaxies  is certainly feasible.
\vglue 6.5truecm
{\tenrm Fig. 3. The numbers of galaxies at each redshift predicted to be
found in I and J selected samples compared to a model fit to the
distribution of the CFRS galaxies (relatively mild luminosity evolution
was assumed - see text).}

\vglue 7.5truecm
{\tenrm Fig. 4. Two examples of spectra of the z $> $ 1 CFRS galaxies. Note
that one is considerably fainter than our nominal survey limit}
\vglue 0.2cm
\noindent
The highest redshift in the CFRS sample is $\sim$ 1.4, owing to the 8600\AA \ts
red limit of the spectra. In order to go to z $\sim$ 2, near-infrared
detectors will have to be used, but fortunately the atmospheric windows
will still allow enough spectral features through to confidently
measure a redshift (figure 5). Based on our CFRS data, $\sim$300
galaxies with 1 $<$ z $<$ 2 are expected in a 5' field at J $<$ 22.
We estimate that their redshifts could be measured in $\sim$4 hours
with a multi-object spectrograph on an 8m telescope.
\vfil
\eject
\vglue 8.5truecm
{\tenrm Fig. 5. A diagram showing the location of atmospheric absorption
(upper panel) and how the wavelengths of
spectral features which are important
in determining redshifts from low resolution data vary with redshift. }
\vglue 0.4cm
\leftline{\twelvebf 5. Conclusions}
\vglue 0.4cm
Our CFRS data shows that the spectra of field galaxies at z $>$ 1 are
very similar to those in the local Universe. They also demonstrate that
surveys of field galaxies up to at least z $\sim$ 2 are feasible with new large
near-IR detectors coupled with efficient multi-object spectrographs.
Such a project will be
relatively easy on 8--10m telescopes.
\vglue 0.6cm
\leftline{\twelvebf 6. Acknowledgements}
\vglue 0.4cm
The CFRS team wishes to thank the CFHT directors and TAC for their
support in the allocation of observing time and in their support of
instrumental innovation.

\vglue 0.6cm
\leftline{\twelvebf 7. References}
\vglue 0.4cm

\medskip
\itemitem{1.} W.V. Schemp, in {\twelveit CCDs in Astronomy, ASP Conf. Series 8}
, ed. G.H. Jacoby, (1990) p111.
\itemitem{2.} A.M. Fowler {\twelveit et al}, in {\twelveit Instrumentation in
Astronomy VIII, SPIE}, {\twelvebf 2198} (1994) 623.
\itemitem{3.} K.-W. Hodapp {\twelveit et al}, in {\twelveit Instrumentation in
Astronomy VIII, SPIE}, {\twelvebf 2198} (1994) 668.
\bye